\documentclass[showpacs,preprintnumbers,amsmath,reprint,aip,jcp]{revtex4-1}
 
\usepackage{bm}
\usepackage{graphicx} 
\usepackage{color}
\usepackage{amsmath}

\begin{document}
 
\title{Universal threshold law for ion-neutral-neutral three-body recombination}

\author{ Jes\'{u}s P\'{e}rez-R\'{i}os}

\affiliation{Department of Physics and Astronomy, 
Purdue University,  47907 West Lafayette, IN, USA}

\author{Chris H. Greene }

\affiliation{Department of Physics and Astronomy, 
Purdue University,  47907 West Lafayette, IN, USA}

\date{\today}

\begin{abstract}

A very recently method  for classical trajectory calculations for 
three-body collision [J. P\'{e}rez-R\'{i}os, S. Ragole, J. Wang and 
C. H. Greene, J. Chem. Phys. {\bf 140}, 044307 (2014)] has been
 applied to describe ion-neutral-neutral ternary processes for low 
 energy collisions: 0.1 mK - 10 mK. As a result, a threshold law 
 for the three-body recombination cross section is obtained and 
 corroborated both, experimentally and numerically. The derived 
 threshold law predicts the formation of weakly bound dimers, with 
 binding energies comparable to the collision energy of the collisional 
 partners. In this low energy range, this analysis predicts that molecular
  ions should dominate over molecular neutrals as the most products 
  formed. 
      
\end{abstract}

\maketitle

\section{Introduction}

A nonradiative collision of two atoms cannot lead to the formation of a stable 
molecule, due to the conservation of energy (it can lead 
to the formation of a transient resonant state). Nevertheless, three
 atoms can collide and eventually, form a molecule {\it i.e.} 
 $A +A +A \rightarrow A_{2} +A$. This is the three-body 
 recombination (TBR) process. TBR processes are one 
 of the main loss mechanisms in systems of ultracold atoms, often 
 dominant in Bose-Einstein condensates.~\cite{Hess-1983,Hess-1984,
 Goey-1986,Esry-1999,Burt-1997,Weber-2003,Fedichev-1996,Suno-2003}
 Recently, the development of hybrid traps technology, where both neutrals 
 and ions overlap in the same spatial region,~\cite{Grier-2009,Zipkes-2010,
 Schmid-2010,Hall-2011,Sullivan-2012,Ravi-2012,Hall-2012} has opened the 
 possibility to study ion-atom interactions and different chemical processes 
 attached to them. These studies have revealed a rich chemistry at 
 such cold temperatures mainly due to charge transfer 
 reactions. However, it has been observed that in high density environments,
  the chemical reactions are dominated by three-body recombination 
 processes.~\cite{Harter-2012} These TBR processes involving neutrals and 
 charged particles have received comparatively little theoretical 
 interpretation.\cite{Francis-2006,Mansbach-1969,Ermolova-2014} 
 In particular, there has been no available prediction concerning 
 the final product states and the dependence of the TBR cross section 
 as a function of the collision energy.

 The present study reports the derivation of a first principles classical threshold law
 for ion-neutral-neutral TBR. This threshold law is numerically tested 
 by comparing its predictions with the numerical results coming from classical trajectory 
 calculations,~\cite{JPR-2014} as well it is experimentally observed in hybrid 
 trap experiments.~\cite{Ulm} The derived threshold law for the TBR cross section 
 shows a power law behavior on the collision. This 
 also has fundamental implications for the final product states for 
 an ion-neutral-neutral, which we demonstrate to be fully dominated by the formation
  of molecular ions instead of neutral molecules.

\section{Results}  
  
For any two-body interaction, the power-law long-range 
tail of the potential establishes a length scale, and associated
 with it an energy scale.These scales define the range where
the collisions exhibit a totally quantal nature. In particular, for
most types of neutral atom-atom interactions, in particular 
for S-wave ground state atoms, the long-range potential is
 dominated by the van der Waals interaction. In that case, 
 the length scale is defined as $2R_{vdW}=(2\mu C_{6}/\hbar^{2})^{1/4}$, 
 and the energy scale as  $E_{vdW}=\hbar^{2}/(2\mu R_{vdW}^{2})$, 
 referred to as the van der Waals length and van der Waals 
 energy, respectively.~\cite{Jones-2006} For alkali-alkali 
 collisions one typically finds $R_{vdW} \sim $ 100 a$_{0}$ 
 and $E_{vdW} \sim$  1 mK. However, for ion-atom 
 interactions, the long-range is dominated by the charge
 induced dipole moment interaction $-C_{4}/r^{4}$, where 
$C_{4}=\alpha/2$, and $\alpha$ is the neutral atom polarizability. 
For this case, we define the polarization length 
$R_{\alpha}=(\mu \alpha/\hbar^{2})^{1/2}$ and the polarization 
energy as $E_{\alpha}=\hbar^{4}/(2\mu^{2} \alpha)$. Typically for 
ion-alkali atom interactions these values are of order 
$R_{\alpha} \sim$ 5000 a$_{0}$ and $E_{\alpha} \sim$ 100 nK, in 
particular for Rb - Rb$^{+}$ and for Ba$^{+}$-Rb are 156 
nK and 104 nK, respectively. Therefore, cold collisions 
(in the milikelvin range) involve many partial waves. For this reason,
 classical trajectory calculations are expected to be reasonably accurate 
 for revealing the reaction dynamics in hybrid trap experiments. In particular, 
 Newtonian mechanics should be applicable for the study of ion-neutral-neutral
  TBR processes.

A recently developed method for the calculation of TBR 
cross sections based on classical trajectories~\cite{JPR-2014} 
has been adapted for the study of ion-atom-atom recombination. 
The method employed relies on mapping the three-body problem 
into a 6-dimensional space (after separating out the trivial center 
of mass motion), where the cross section emerges as a generalization
 of  the well-known two-body cross section.~\cite{JPR-2014} Hyperspherical 
 coordinates~\cite{Avery} are used for representing the 
 positions and momentum vectors in the 6-dimensional 
 space leading to a very efficient sampling of the phase-space. 
 In hybrid trap experiments, the kinetic energy of 
 the ion is almost two orders of magnitude higher than the energy of the
 ultracold neutral atoms. Keeping this in mind, the 
 classical trajectory calculations (CTC) presented here have been 
  performed by fixing one of the hyperangles associated
 with the initial momentum, guaranteeing that 95 \% of the collision 
 energy is associated with the kinetic motion along the direction 
 of the ion. However, general CTC without any constraint have 
 been also performed and they will denoted as FCTC in the present work.

Figure 1 displays our results for the TBR cross section for
$^{87}$Rb$^{+}$ -$^{87}$Rb -$^{87}$Rb, and for
$^{138}$Ba$^{+}$ -$^{87}$Rb -$^{87}$Rb, which are presented in panels (a) and 
(b), respectively. The TBR cross section for the systems at hand 
has been computed by running $10^{5}$ trajectories per collision 
energy. During the simulation the energy is conserved up to the 
fifth decimal place, and the same is observed for the angular momentum. 
 Details about the numerical method employed to solve 
 Hamilton's equations of motion, in conjunction with the sampling
 of the initial conditions, can be found elsewhere.~\cite{JPR-2014} All of 
 the calculations have assumed that neutral-neutral interactions, as well 
as ion-neutral interactions occur along one single potential energy curve. 
Concretely, Rb - Rb collisions are assumed to occur through the $^{3}\Sigma$ 
 potential, {\it i.e.}, the spin flip transitions have been neglected. In particular, 
 the potential of Strauss {\it et al.} has been employed.~\cite{Strauss-2010} On 
the other hand, the ion-neutral potential is described by the model potential
 $-C_{4}(1-(r_{m}/r)^{4}/2)/r^{4}$, where $C_{4}$ denotes 
the experimental long-range coefficient of the interaction, which is taken as 
C$_{4}$=160 a.u. along this work, and $r_{m}$ 
represents the position of the minimum of the potential. For Ba$^{+}$ - Rb 
interaction the value of $r_{m}$ is taken from the work of Krych 
{\it et al}.~\cite{Krych-2011}, whereas for Rb$^{+}$ - Rb the same 
magnitude comes from Jraij {\it et al}.~\cite{Allouche}

Fig. 1 shows that the TBR cross section for 
the systems depends quite smoothly  on 
the collision energy, which is independent of the nature
 of the system at hand; this suggests an
emerging universal threshold behavior. Figure 1 (b) demonstrates
that the constraint in the hyperangles associated 
with the momentum (CTC) does not affect the general trend 
of the TBR cross section, as good agreement is seen with the FCTC results. 
This suggests that the emerging threshold behavior is largely independent 
of the initial momentum vectors of the system. 
In other words, the dependence of the TBR cross section must be primarily
controlled by the interatomic potential among the involved particles. In order to 
emphasize the dependence of the TBR cross section on the collision 
energy, the power law fits of the CTC results are shown as dashed lines, and the 
fitting parameters are presented in Table I.  
 
  \begin{figure}[t]
\centering
 \includegraphics[width=9.5 cm]{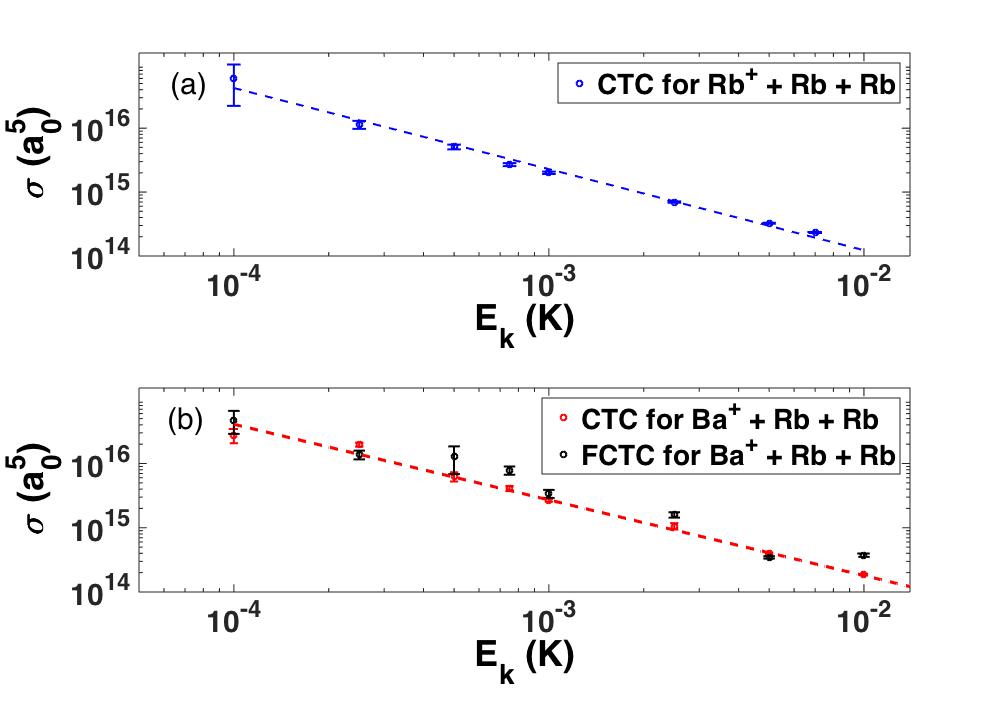}
 
 \caption{(color online). Three-body recombination cross section 
 (in a$_{0}^{5}$) as a function of the collision energy (in K). Panel (a) 
 $^{87}$Rb$^{+}$ - $^{87}$Rb - $^{87}$Rb ; the circles 
 represent the numerical results by means of CTC whereas the dashed
  line stands for the fitting of the points. Panel (b) 
  $^{87}$Rb$^{+}$ - $^{87}$Rb - $^{87}$Rb ; red circles
  represent the numerical results by means of CTC, the black circles denote 
  the results using CTC without any constrain on the (see text for details), the
   dashed line stands for the fitting of the obtained CTC results. The fitting function 
   assumed for both systems is $\sigma(E_{k})=\gamma E_{k}^{\beta}$, and the 
   results are presented in Table I.}
 \end{figure}

For low energy collisions, particles probe
the long-range tail of the interaction during most of 
the collision time. The case of ion-atom-atom 
collisions is not an exception, but it requires 
some special consideration in the three-body 
recombination system of interest here, since there are two
 different long-range potentials involved in the problem, 
 namely atom-atom and ion-atom. From a long-range 
 perspective, the ion-atom interaction is of course far 
 more attractive than the atom-atom interaction. For any
orientation, energy and impact parameter, the 
collision begins when the trajectories of the atoms
 start to deviate from uniform rectilinear motion.
 This occurs when the interaction potential where the
 moving atoms are currently located is comparable to the collision
 energy, {\it i.e}, $E_{k} \approx C_{4}/r^{4}$. Here, we will 
 assume that this value of the radius defines the maximum
 impact parameter associated to a TBR event, {\it i.e.} 
 $b_{max}(E_{k}) = \left( C_{4}/E_{k} \right)^{1/4}$. In the 
 present formalism the TBR cross section is defined as~\cite{JPR-2014} 

\begin{equation}
\label{eq-1}
\sigma(E_{k}) \propto \int_{0}^{b(E_{k})_{max}}b^{4}db,
\end{equation}
 
 \noindent 
where a unit opacity function has been assumed. Finally, taking into 
account the expression for $b_{max}(E_{k})$ and substituting it in 
Eq.~\ref{eq-1}, $\sigma(E_{k})\propto E_{k}^{-5/4}$ it is found. This 
result is compared in Fig.1 with a fit of the CTC numerical results shown in Table I. 
In particular, the fitting function employed has the form 
$\sigma(E_{k})=\gamma E_{k}^{\beta}$, where $\beta$ is related with the
 energy scaling law associated with the TBR cross section at low 
 energies, which reveals the underlying classical threshold law. The errors 
 reported in Table I are associated with a confidence interval of 95 \%.

\begin{table}[h]
\caption{Classical threshold law for the TBR cross section. 
A power law dependence of the TBR cross section as a function 
of the collision energy is assumed and used as a fitting function 
for the CTC numerical results presented in Fig. 1. The error 
on the fitting parameters are associated with a confidence 
interval of 95 \%.}
\begin{tabular}{ c c c }
\hline
  System & $ \gamma$ (a$_{0}^{5}$)&$\beta$ (dimensionless) \\ 
  \hline
   $^{87}$Rb$^{+}$ - $^{87}$Rb - $^{87}$Rb & (7.94 $\pm$ 2.72) 10$^{11}$ & -1.178 $\pm$ 0.068 \\
   $^{138}$Ba$^{+}$ - $^{87}$Rb - $^{87}$Rb &  (3.57 $\pm$ 0.07) 10$^{11}$  & -1.269 $\pm$ 0.132  \\
   Classical threshold law & & -1.25 \\ \hline \hline
\end{tabular}
\end{table}

In Table I, it is observed that energy scaling law for the TBR cross section 
numerically obtained for both systems, $^{87}$Rb$^{+}$ - $^{87}$Rb - $^{87}$Rb and 
 $^{138}$Ba$^{+}$ - $^{87}$Rb - $^{87}$Rb, are in a good agreement
  with the predicted energy scaling law associated with the derived threshold law.
This suggests that the derived threshold law is satisfied in different systems 
under different dynamical conditions. A similar threshold law 
was derived and numerically confirmed for 
atom-atom-atom collisions.~\cite{JPR-2014} On the other hand, the 
presented results for  $^{138}$Ba$^{+}$ - $^{87}$Rb - $^{87}$Rb 
have been corroborated in hybrid trap experiments, yielding a 
first principles explanation of the observed ion losses caused by TBR processes.~\cite{Ulm}
 
 The derived threshold analysis has implications beyond the energy dependence 
 of the TBR cross section. It predicts that the final 
 molecular product states will be weakly bound. This is due to the correlation 
 between the collision energy and to the binding energy, which has been observed 
 in previous calculations~\cite{JPR-2014} for neutral-neutral-neutral TBR processes, 
 and confirmed in the present calculations for ion-neutral-neutral TBR. Moreover, 
 the CTC and FCTC calculations both establish that molecular ions 
 constitute the dominant product channel. This may be associated with the
  dominance of the of the ion-neutral long-range interaction on the cross section.

 \section{Conclusions}

 A classical threshold law for ion-neutral-neutral TBR processes has been 
 derived, numerically confirmed, and also experimentally corroborated. The 
 present threshold behavior apart from the prediction of the energy dependence of 
 the TBR cross section, has significant implications in terms of the expected reaction 
 products. In particular, the low energy threshold law for TBR involving one 
charged particle and two neutrals predicts that 
predominantly molecular ions are formed as products of 
the TBR. This might seem counterintuitive, since in principle one 
 expects some influence of the atom-atom 
 interaction on the dynamics. However, as is discussed above, the ion-atom interaction
 presents a more attractive nature than the 
 atom-atom, and hence it dominates the 
 dynamics of the collision. Nevertheless, it has been 
 checked that the situation changes as the collision
 energy increases. In that case, the atom-atom 
 interaction eventually becomes as important as the ion-atom 
 interaction. 

The observation of molecular ions in hybrid traps will be quite challenging due to 
the presence of electric fields. In particular, weakly bound molecular ions 
are easily dissociated in the the Paul trap by the external electric field. For instance, with an 
 electric field $\sim$ 1V/cm the molecular ion will dissociate 
 on a time scale $\sim$ 100 $n$s (the time scale of the vibrational period). Sorting out the 
 three-body reaction is also challenging because many of  the reactions occur in the presence of 
 the dipole trap light or MOT light. This light can couple the very weakly-bound 
 vibrational states with highly excited states of the molecular ion, leading to far 
 richer chemical processes involving excited states.

\section{Acknowledgements}

This work was supported by the Department of Energy, Office 
of Science, under Award Number DE-SC0010545. The authors acknowledge
Francis Robicheaux, Johannes Hecker Denschlag and Artjom Kr\"{u}kow for 
many fruitful discussions.

\bibliography{TBR}

\end{document}